# Text Classification for Task-based Source Code Related Questions

**Sairamvinay Vijayaraghavan**  **Jinxiao Song**

**David Tomassi**  **Siddhartha Punj**  **Jailan Sabet**

## 1 Introduction

Classification of text is an important field of research and a core task in Natural Language Processing (NLP). It spans many different domains from determining "fake" news, finding spam emails, and language detection. A common problem in software development is generating the appropriate code snippet for a task. There are communities for software development that allow people to ask questions and provide answers such as Stack Overflow. These answers in Stack Overflow represent snippets in a few lines which contain mostly API Calls from various libraries in Python. The issue that arises is that many people will provide completely different answers for a particular question in which many cases the answers are wrong. We pose the task of determining if a pair of a question/problem and a corresponding code snippet is appropriate or not as a binary classification problem.

Our goal is to learn from these pairs the correct code snippets for certain problems. An example of this is the code snippet,

```
shutil.copy('file.txt', 'file2.txt')
```

and the question, "how to copy one file's contents to another in python?". We will make use of deep learning based sequence-to-sequence models to learn from a code corpus and corresponding intent converted into 'yes' or 'no' labelled data. The creation of the 'yes/no' questions will be conducted by using the existent intent-snippet pairs as positive and then create artificial negative intent-snippet pairs by pairing intents with snippets which belong to a totally different intent-snippet sample pair which does not even exist. We will have the positive "yes" questions as the standard answer to the question. The negative "no" questions will be the artificial intent-snippet pairs which we create by random sampling of the snippets with the intents ensuring they don't belong to the same question. In order to provide an additional development vocabulary for training the code corpus, we would be using a development vocabulary from outside corpora for capturing the finer semantic meanings and distribution across the code tokens within the Python. So, we plan to ensure that a balanced set of positive and negative samples for training our models for classification. Our code repository can be found below.[1]

## 2 Related Work

Source code generation has wide applications including automated program repair where the goal is to generate correct programs from buggy ones [7] [9], source code translation where source code is the input and the output will be source code in a different programming language [13][6] In addition, code generation is highly applicable to users that may want to complete a specific task that they can formulate in natural language, but are unable to mimic since they don't have prior background of a language or they are simply having trouble finding particular library functions in a language to complete a given task[19]. Some of the previous ways to generate code from natural language include using transition based neural semantic parsers to map natural language utterances into formal meaning representations. This is done by creating a transition system with rules to construct an AST given the natural language input, and then using the AST as an intermediate meaning representation as a guideline to follow to convert the original input back into code[18].

More importantly, we were motivated strongly to pursue this task because of the CoNaLa base-line model established by the authors[17]. The authors have trained a logistic regression based classifier for predicting whether the snippet provided matches the intent and it is doing a correct task. The model predicted labels -1 for no and +1 for yes. So, this was our primary motivation for coming up with a deep learning model for fine tuning the classification task in addition to a generic seq2seq model for code generation. We are planning to use the same methods to induce artificial false examples as in the paper.

---

[1] https://github.com/Sairamvinay/Code-Generation-Classification-QA

In terms of parsing the code snippet, there are not many great resources available online to suit our special requirements while handling the vocabulary of the code snippets corpus. However, the paper[10], which presents a technique for inferring the dependencies needed to execute a Python code snippet without import error, did give some inspiration on the possibility of using methodology to detect API's and function name.

There have been many different approaches to represent source code for learning word vectors. Code-related NLP tasks require the breakdown of source code into a form that can be utilized by machine learning algorithms, predominantly as vectors. However, due to the structure of code, the same technique used for natural language cannot be applied to code (i.e. linear break down, word by word); instead, code is often broken down into abstract syntax trees, which then can be converted into code vectors. Bilgin et al.[4] implemented this idea to create a machine that detects vulnerabilities in code with machine learning algorithms. However, having to implement this step for every code-related NLP task would be time consuming. Instead, numerous authors addressed this concern by creating an open-source implementation. For instance, DeFreez et al.[8] proposed func2vec, an algorithm that maps functions of C source code to vectors in a vector space grouped by similarity using static program paths. Alon et al. (2018)[1] introduced a now popular framework called code2vec that creates code vectors ("code embeddings") for Java source code, utilizing this framework to create a model for predicting method names based on vector similarities. Code2vec has been utilized in several NLP works, including Briem et al.[5], in which a machine took in code embeddings created by code2vec to detect off-by-one errors using binary classification. Arumugam [2]adapted code2vec for the CodeSearch-Net challenge[11]; in other words, semantic code search that, when given a query, can retrieve relevant code. Code2vec also inspired other implementations, such as PathMiner, now called astminer, developed by Kovalenko et al.[12] to create code embeddings for Python source code using code2vec's algorithm.

## 3 Dataset

For validating our hypothesis, we chose the Code/Natural Language Challenge (CoNaLa) dataset from CMU[17] and the StaQC dataset[16] from StackOverflow in addition. We chose StaQC dataset since we found out that CoNaLa was having too few examples after data de-duplication using Question ID. The CoNaLa dataset contains around 598k samples in form of Json lines formatted files automatically data mined from Stack Overflow which contains the question ID, intent of the code snippet presented, code snippet, parent answer post ID, and an unique ID for the post. In addition to the above set, there is another around 2K samples of manual curated data which contains question ID, intent, rewritten intent and also the code snippet. We combined both of these datasets together in order to get all the CoNaLa examples.

The StaQC dataset contains over 147K python based intent-snippet samples with a similar format to CoNaLa, containing a code snippet, intent, and a labelled question ID for the Stack Overflow post the intent and snippet refer to. However, it does not contain rewritten intents as in CoNaLa that incorporate variable names and function arguments back into the intent for a better representation of the task that the code snippet solves. For our tasks, we particularly analyzed the single code answer posts (around 85K samples) from StaQC, and we filtered samples by analyzing code snippets by applying a threshold of a size of 5 lines or less on the code sequence length for analyzing just short snippets. We conclude that the combined dataset of both CoNaLa and StaQC would provide all the snippets from the Python3 programming language (more on this in the next section).

## 4 Methodology

### 4.1 Dataset Preprocessing and Curation

For the data curation, we initially analyzed the CoNaLa dataset alone, which consists of two main data files that we plan on combining in order to evaluate our tasks. We had chosen to inspect mainly the files named CoNaLa-train and the CoNaLa-mined. Both of these files contain key value pairs of question ids, intents, and the code snippets associated with the question.

We found out there were too many duplicate intent-snippet pairs within the dataset by inspecting the question ID across these files. So we performed data de-duplication on the CoNaLa dataset. However after data de-duplication (described more in depth below) we were only able to retain 4302 unique intent-snippet pairs. In order to increase the size of the training corpus, we combined the CoNaLa dataset with the existent StaQC dataset consisting of intent-snippet pairs. After combining these datasets, we had obtained around 43,000 pairs of unique intent-snippets. We describe our pre-processing steps in the CoNaLa section below

### 4.1.1 CoNaLa

CoNaLa-train and the automatically mined CoNaLa-mined data files differ from each other based on the features present in every sample. CoNaLa-mined dataset contains 600k examples from Stack Overflow, probabilities given by the baseline model (discussed in the related works) that a snippet is a correct answer for a given intent. We retained the intent, code snippet, and question ID for validating de-duplication.

We parsed CoNaLa-train for key value pairs of question ids, and a list of code snippets they're associated with. In order to remove snippets that are too similar to each other or resemble the same answer (duplicates are existent in the dataset), we calculated the cosine similarity of code snippets associated with a particular question and using our own vocabulary file of code tokens that are used in the Python language. We had removed those samples which have very similar snippets for the same question.

In order to improve our approach in identifying similar snippets, we used the vanilla cosine similarity method in particular to evaluate how similar each code snippet is to another. In order to establish whether an answer was too similar to another, we established a similarity threshold of at least 0.5 (that is 50% similarity or above means it will be removed and termed as a very similar code snippet). If a snippet passes this threshold in comparison to another snippet based on the similarity matrix, then it will be removed. We also noticed while working on CoNaLa-mined, that most of the answers listed (we refer to the code snippets listed for a single intent provided across different samples) were not at all similar to each other and also were not relevant in the context of the problem. For example, the question being, "Sort a nested list by two elements" and a corresponding answer was (-10, 'Anthony'), which was absolutely irrelevant. In order to resolve this peculiar problem, we had used the results from the logistic regression baseline developed [17] for prediction of the very same task. We went ahead establishing a probability threshold for this logistic regression baseline results listed in the CoNaLa-mined dataset. We performed some statistical analysis on that likelihood of the answer being valid and found an average likelihood for the best answer per question being 23 % with a standard deviation of 15 %. After a manual inspection of answers within a standard deviation of the mean, we have empirically chosen a probability threshold of at least 0.5 (believing 50 % probability and above ensuring a valid answer for the task presented) to choose answers that were accurate and relevant to the question asked. This is highlighted in fig. 1.

Additionally, we inspected the train and test sets from CoNaLa to verify if the snippets in these sets are overlapping within the CoNaLa-mined (golden standard) set. In such cases, we removed these overlapped samples and retained just the non-duplicate/new samples. The same previous approach with the similarity metric for filtering similar answers across different samples was applied. Also, we determined question similarity across the different sets to determine whether there was any overlap between the questions referred. After we completed this task, we combined the cleaned up datasets of CoNaLa-train and CoNaLa-mined into a single dataset of questions and their possible answers by reformatting CoNaLa-mined into intent and snippet(s) segments.

In the final dataset, it consisted of 4,302 possible unique intent-snippet samples in our curated CoNaLa dataset.

### 4.1.2 StaQC

After we found that only very few samples remained after removing duplicate answers to questions and overlapping questions in the CoNaLa-train and CoNaLa-mined datasets, we utilized the StaQC dataset for providing further examples to train. We cleaned the raw StaQC dataset of Stack

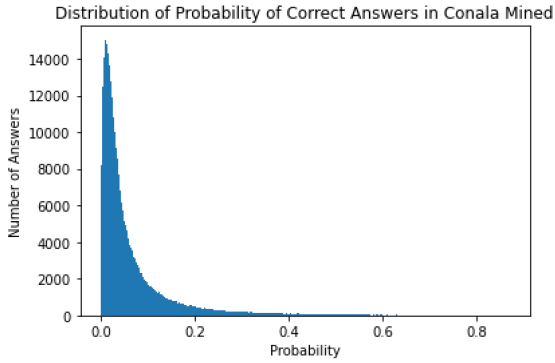

Figure 1: Distribution of probability of correct answers established by Logistic Regression baseline in CoNaLa mined file

Table 1: Question Answer Dataset

|        | Positive | Negative | Total  |
|--------|----------|----------|--------|
| Conala | 4,302    | X        | 4,302  |
| StaQC  | 39,817   | X        | 39,817 |
| Total  | 43,238   | 41,922   | 85,160 |

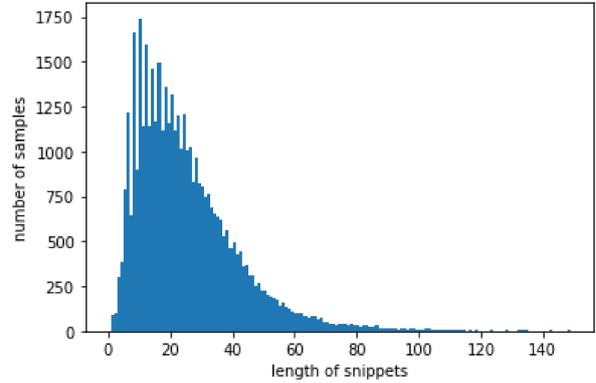

Figure 2: Code snippet character length distribution: 0-150

Overflow questions by filtering just the single code answer posts in Python and later combined it with the CoNaLa dataset files.

We cleaned up the code snippets in particular by removing unnecessary punctuation/spaces, one line python comments with hashtags, multi-line comments, as well as code that contained Python interpreter symbols by themselves and with value return lines. As mentioned previously, we kept a maximum threshold size of 5 code snippet lines for easier analysis of the short code snippets by our code parser. Figure 2 shows the code snippet character length distribution. We found our largest snippet length (number of code tokens) to be 635, but snippets beyond the length of 150 are not shown because there are very few of the samples (in fact negligible given the number of samples present in the corpus). We end up with 39,817 positive samples in our curated StaQC dataset.

### 4.1.3 Combined

We combine both the curated CoNaLa (4302 unique samples) and StaQC (39817 unique samples) datasets. We ensure that overlapping question answer pairs are not duplicated in the final dataset. Also, we found that some of the snippets were wrongly formatted when they were mined using the Stack Overflow API. These samples were not even recognized when we parsed using the 'tokenize' library in python for tokenize the Python code. So, We had removed these faulty and duplicate samples (881 samples from CoNaLa and StaQC to be precise), as shown in table 1.. Finally, the dataset consisted of 43,238 positive samples.

### 4.1.4 Negative sample creation

For training binary classifier, we added a class variable of 1 to all matched intent-snippet pairs to determine a positive sample. To create negative samples, we iterate through each intent and randomly sample a code snippet from the whole dataset (which obviously does not belong to the same question ID of the intent). In other words, we take each intent and pair it with a wrong code snippet. These negative samples will be assigned a class variable of 0. We end up with 41,922 negative samples.

## 4.2 Parsing the inputs

### 4.2.1 Intent

For parsing the intents, we did not have to do much of complex pre processing techniques. The model needed to learn "what" task needs to be solved and generate the code sequence appropriately. So, in order to solve this issue, we had cleaned the natural language based intents by removing the basic punctuation characters such as ?,!,:,",;, and - and also had converted each character into lower case. We had retained the dots (. operator) since some of the intents contained method calls which were in direct relation to the code sequence we were predicting. We had used intents for the negative samples data generation also and hence we had used this cleaned intents for our further tasks.

### 4.2.2 Code Snippet

In order to support the learning the vocabulary by the model for analyzing the code tokens present in the Python, we want to utilize a large Python3 based vocabulary corpus based on outer corpora. This larger vocabulary contains various programming language tokens present in the Python programming language.

Initially, we decided to just use the code snippets from the CoNaLa dataset. However, for developing the vocabulary and develop our code based embeddings, we pre-process an external dataset, CodeSearchNet which provides a much larger code corpus to expand our code vocabulary base. This dataset contains code in the form of user defined functions and the code refers to the functions crawled from GitHub repositories and also various Stack Overflow questions. We will separate the code into individual tokens and store the unique tokens as a list of vocabulary. For basic tokenizing, we would use the built-in library 'tokenize' which would separate the snippet into its individual code tokens.

**API calls retrieval** We encountered multiple issues while parsing the Python code in the CoNaLa dataset. Since this dataset includes short snippets of code, there are mainly API calls from numerous libraries in Python and also various methods of the data types (such as lists, strings, dictionaries etc.) in Python. So, we have to train the model ensuring that it learns in understanding the intricacies of the language by identifying the user-defined variable/function names and also retaining the library names and the method names associated with API calls.

While we are working on distinguishing the user-defined names and the existent API call based names, we have arrived at a particular proposition. In order to build a robust parser for generating the vocabulary, we need to add additional capabilities for the code parser to identify the user-defined names and also retain the various API calls of methods from Python programming language. For those terms, the parser should not mark those as names that can be interchangeable (that is map these names (or also called as normalizing these tokens) to a common token "<VAR_NAME>"). Our approach is using a manual approach for helping the parser to detect API based calls of certain most commonly used libraries in Python and also the methods of various data types in Python.

```
INTENT: Fastest way to initialize numpy array with values given by function
Initial CODE

np.fromfunction(f, shape=(d1, d2))

PARSER's OUTPUT
[
    "np",
    ".",
    "fromfunction",
    "(",
    "<VAR_NAME>",
    ",",
    "shape",
    "=",
    "(",
    "<VAR_NAME>",
    ",",
    "<VAR_NAME>",
    ")",
    ")"
]
All variable/function names to normalize:
['d1', 'd2', 'f']
```

Figure 3: The final code parser output for a sample snippet

We incorporate the most popular Python-based Library names and the associated methods for helping in fine-tuning the code parsing for providing a much more closer to a real-life scenario for Python language code to understand which names are interchangeable and which are not.

In order to analyze the most frequent calls for collecting the manual API based methods, we used different strategies for each corpus (the current combined CoNaLa and the StaQC corpus and the development CodeSearchNet corpus). We plan to find the top 40 API/data types calls for each of the corpora and then finalize on manually collecting the API method calls by selecting the most frequent API names in each corpora from both. We narrow down the list by also keeping the intersection between the API names we find from each corpora.

**CodeSearchNet based parsing** For CodeSearchNet, since each code snippet is stored as a Json object and it was referring to GitHub repository URL links to the source code, we built a web scraper that can visit the GitHub repository URL link associated with the sample and scrape just the import commands. There are 21 000 URLs that the web scraper needs to scrape and store the frequency count of all those library names in order to help to list the top 20 API libraries that are being used.

**Current Corpus based parsing** For the combined dataset between CoNaLa and StaQC, we had used the intents for finding the most commonly referred APIs. The reason was because we inspected that APIs are referred commonly in the intents and the code snippets were rather one or two liners which didn't bear any import statements. So, we had extensively cleaned and filtered the intents alone for the API calls. From the intents, had removed the most commonly occurring natural language words called stop words in English by applying a filter on the stop words and then using PorterStemmer from NLTK library for stemming the words into its root forms. Then we had ended up with a vocabulary of the API names and/or the data types associated with the problem task. We then find the list of the top most frequently occurring words within these intents and these represent the most frequently occurring API/data type calls used across this corpus fig. 4.

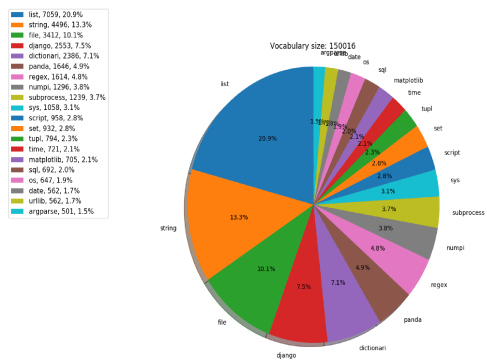

Figure 4: CoNaLa API Distribution

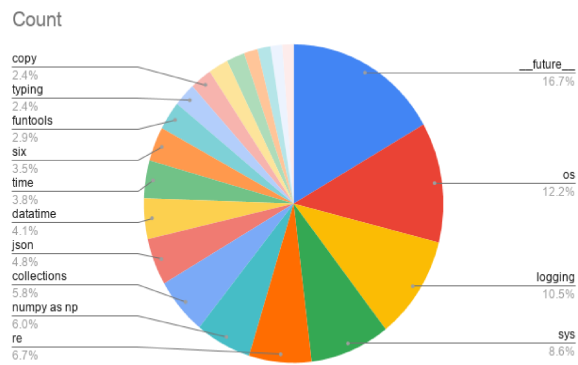

Figure 5: CSN API Distribution

Finally, we list top 40 API libraries in a file with its function and method names and then identify just the user defined names. This helps us making our vocabulary robust and more generalized.

With the top 40 API libraries, we can identify which is the variable name that is user defined. Then, we decided to normalize the user-defined names for functions and variables as a single normalized token <VAR_NAME> which would indicate that this token can be replaced by any user-defined name present across the corpus (a flexible interchangeable variable name). We use these normalized and tokenized snippet sequences for developing our embeddings.

## 4.3 Embeddings

### 4.3.1 Intent Embeddings

For the encoder based LSTM model described below, we had planned to use an Embedding layer for the intents which would be pre-trained. For this, we intend to use a very simple Skip-gram based Word2Vec model trained on the current corpus of the cleaned intents. We used a fixed set of parameters such as window size as 4, and trained it for 15 epochs. We used a minimum count of 1 which ensured that words with count less than or equal to 1 are excluded in the training process.

### 4.3.2 Code Embeddings

Similar to the use of word embeddings for natural language analysis, breaking code snippets down into a continuous distributed vector representation (i.e. creating "code embeddings") is necessary for creating a vocabulary for Python and accomplishing our task [2]. While there are many methods that produce code embeddings using abstract syntax trees [2][12][1], these code embeddings are created for Java source code rather than Python source code. In addition, Azcona et al. compared numerous methods of creating code embeddings, including AST and word tokens, and found that word tokenization of code leads to F1 score slightly higher than when using ASTs as code representation (57.93% compared to 56.91%, respectively) [3].

We decided to use pre-trained word embeddings for our source code to help us in experimenting our tasks. We felt that pre-trained embeddings would help in understanding the semantic meanings of code tokens across a corpus. We wanted to experiment with different type of embeddings for our code snippets. Therefore, we utilized two methods to create vector representation of code often reserved for word representation: Word2Vec [14] and GloVe [15]. In addition, we trained these embeddings on two different corpora: the current dataset of the CoNaLa and StaQC (the current corpus) and a much larger dataset with the corpus CodeSeachNet added to the current dataset (we refer to it as the "CSN corpus"). This resulted in four

Table 2: Vocabulary size for each embedding

| Embedding | Vocabulary size |
|---|---|
| W2V on Current | 34669 |
| GloVe on Current | 39106 |
| W2V on Current + CSN | 22814 |
| GloVe on Current + CSN | 58434 |

Table 3: Top Ten Frequent tokens vocabulary count for each corpus used

| Token Found | Current | Current + CSN |
|---|---|---|
| <VAR_NAME> | 232,712 | 1,186,319,608 |
| ( | 94493 | 323,427,063 |
| ) | 94403 | 323,330,993 |
| . | 76500 | 331,126,158 |
| , | 82165 | 281,027,191 |
| = | 51499 | 253,933,989 |
| : | 29317 | 171,159,189 |
| [ | 33017 | 146,598,615 |
| ] | 32987 | 146,575,980 |
| in | 12719 | 35,940,949 |
| if | 4562 | 69,695,925 |

types of word embeddings as in table 2

For each of the models, we fixed on a common set of hyper parameters which included a context window size of 15 words, a token minimum count of 0 (with the exception of the CSN word2vec model, which had a minimum count of 5 due to its large size). The models were each trained for 10 epochs, and we chose mainly the skip-gram implementation for the word2vec models.

Following the creation of these models, we performed an analysis to gain a better understanding of the similarities and differences of each. For instance, each model tended to have similar frequent tokens. As observed in the frequency tables of the code tokens vocabulary, which can be seen in table 3, the ten most frequent tokens had been the same all across except for the last token (the token "if" for the CSN dataset compared to the token "in" for the current dataset). We also projected all of the tokens of each dataset, as well as the ten most frequent tokens for each dataset, to a Principal Component Analysis space for better insight into the mappings of each token. The results of these can be seen in section 6.3.

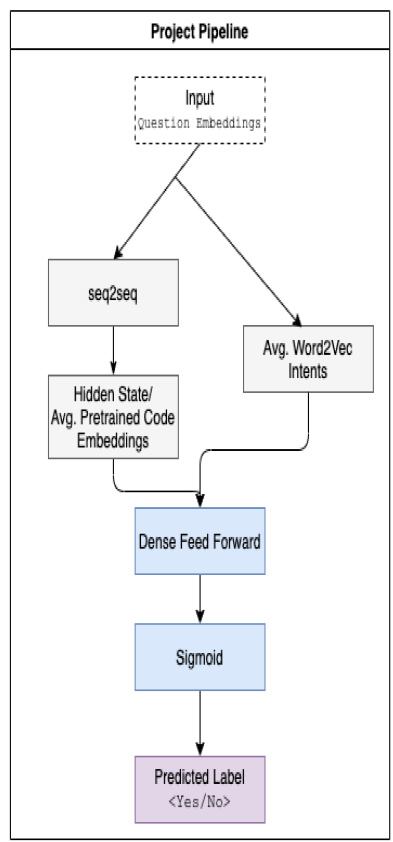

Figure 6: The overall pipeline of our project

## 5 Modeling

### 5.1 Model Description

It was a two part architecture of our model structure. First we train a simple seq2seq model for predicting the code sequence and also the hidden state vectors. We train only the positive samples of the intent-snippet pairs for the seq2seq model. We use an encoder-decoder LSTM based model which was described in the below section.

Then, we use a binary classification model which uses two inputs: intent and code based embeddings described below and then this model was used to predict the final label of the intent-snippet pair, which will be predicted as a probability of the sample pair being a positive (or it is the appropriate task). Our pipeline is described in the figure as fig. 6

### 5.2 Training information

#### 5.2.1 Seq2Seq

For the first model of the project, we used a seq2seq model. The seq2seq architecture is an encoder-decoder architecture that consists of two LSTM networks: the encoder LSTM and the de-coder LSTM. The input to the encoder LSTM is the natural language question utterance and a code snippet as a sequence of code-based tokens. The input to the decoder LSTM is the code snippet pre-pended with a start-of-sentence (<START>) token. The output is the actual target code snippet with the end-of-sentence token (<END>) appended at the end. For the decoder LSTM especially, we need to generate two versions of the code snippets: one with the start-of-sentence token pre-pended and the other with the end-of-sentence token appended at the end.

Then, we applied the tokenization and padding technique on the intents and snippets separately. Tokenization split a sentence into the corresponding words for intent. For the snippets, we get it into a sequence of code tokens. We then convert these words to integers based on the increasing vocabulary count in its corpus. We curtail the vocabulary size to the top frequent 5000 words for the best efficient results. Also almost 80% of the code tokens in the vocabulary were having a frequency of less than 5 across the current corpus we are working on.

For the padding, we need to fix code snippets as the input and output decoder to the same length which we fixed as 50 again due to the same reason that longer snippets were very rare in the corpus. Similarly we fixed the maximum length of intent sequences as 35.

For each of the LSTM layers, we had used embeddings with the pre-trained embeddings for both the intent and the code based sequences which will be used to feed into the LSTM layers (both encoder and decoder). For intent, we have to load W2V vectors from the pre-trained intent embeddings to represent each intent token (word) for encoder LSTM input. Similarly, for code snippets, we have tried the different embedding techniques on w2v and glove between current-corpus and combination of current-corpus with development corpus CodeSearchNet just as in table 2.

In the entire model, we decided to have 100 units per layer with activation of hyperbolic Tangent activation function for both encoder and decoder LSTM and a recurring sigmoid activation for each of the LSTMs. For the encoder LSTM, the models takes in a sequence of words representing a single intent sentence of fixed length of 35. The embedding layer was freezed since the embeddings were already pre-trained.

Similarly, the decoder LSTM takes in an input snippet sequence which has the code sequence of length 51 tokens and the <START> token pre-pended at the beginning of the code sequence. The decoder LSTM used the last hidden state and cell state from the encoder and the input sentence, which actually became the output target sentence with an <END> token appended at the beginning. We predict the next token at every time step and it returns the probabilities for each word in the code vocabulary. We pick the next word which will be the most probable word of the sequence. We will be providing the one-hot-encoding of the target code sequence for each token within every sample and this is used for the training purpose. We had to predict the next word at every time-step. So, the final layer of the model is one fully connected layer which uses the softmax activation function.

### 5.2.2 Binary classifier

The second part of our model, the binary classifier, takes in two inputs: an intent based embedding created (with Word2Vec), as well as the averaged embedding from the seq2seq model outputs for the code sequence. The average embedding for the code is calculated in two ways. One method used the element-wise hadamard product of the hidden state vector and the context state vector and another method was the average of the sum of all the pre-trained embedding vectors for each of the predicted code token in the code token sequence predicted. There are 4 different types of embeddings used for this experiment, as mentioned previously in table 2. Both of these code based embeddings were ensured to provide a vector of fixed dimension, which which are set to a constant size of 100. So, this was because we fixed the hidden layer size for the decoder LSTM as 100 units.

The binary classifier structure is as follows: 2 Input layers are created to take in the intent and snippet embeddings, which are of size 100 as previously mentioned. These two vectors are then concatenated into one single long vector as the input for the model internally. We establish this by concatenating the Input Layers in Keras. Then, in order to consider the context of the question as well as the generated code sequence output from seq2seq together, we concatenate both of them. For the rest of the model, we use 3 fully connected hidden layers of decreasing size (100-50-25) alternating with dropout layers with ReLU activation functions. We choose our dropout rate as 0.5 after each of these hidden layers to match the decreasing size of the hidden layers. After these, we have a final dense hidden layer of size 1 with a sigmoid activation function to classify whether our intent input and generated code sequence from seq2seq are matched.

Table 4: Seq2Seq Test Accuracy and Loss for models.

| Seq2Seq Model | Loss | Accuracy (%) |
|---|---|---|
| current GloVe | 0.96 | 78.33 |
| current Word2Vec | 1.05 | 76.95 |
| current GloVe + CSN | 0.98 | 78.00 |
| current Word2Vec + CSN | 1.13 | 75.34 |
| Hidden State Vectors | 1.30 | 76.71 |

We compile the above model using binary cross entropy as a loss function, and Adam as an optimizer. For the metrics, we only used accuracy for the seq2seq model but for the binary classifier, we used accuracy, F1-scores, Area Under Curve for Receiver Operating Characteristic curves, and Precision-Recall curves. For both the models, we used training and validation sets. After padding both the training and validation intents and training snippets used, we obtain the intent based average embedding into a word2vec embedding, and the code snippet embedding using one of the unique embeddings or the hidden state vectors from the seq2seq model mentioned above. We trained for 25 epochs and batch size was set as 256.

For evaluation, we evaluated 1000 test samples and while evaluating, we perform the inference phase of the encoder-decoder model for predicting the code sequence for every sample from which we obtain the average embedding for the code sequence. The test sample distribution was found as follows: 515 are positive and 485 are negative samples.

For the final results, we will provide Receiver Operating Curves and Precision Recall curves and the associated Area under Curve (AUC) scores for each of these curves. In addition to the same, we plan to report F1-scores and also the accuracy scores for the overall classification task. In order to gauge efficiency of the seq2seq as well as the classifier model, loss training plots will also be provided.

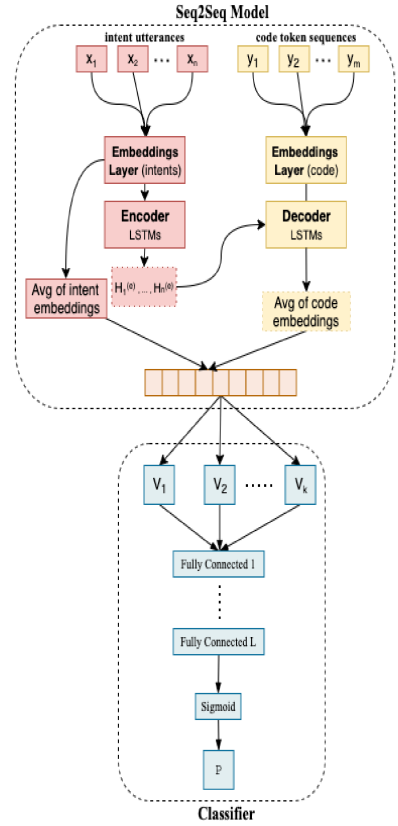

Figure 7: The baseline model based of the code embedding of our project

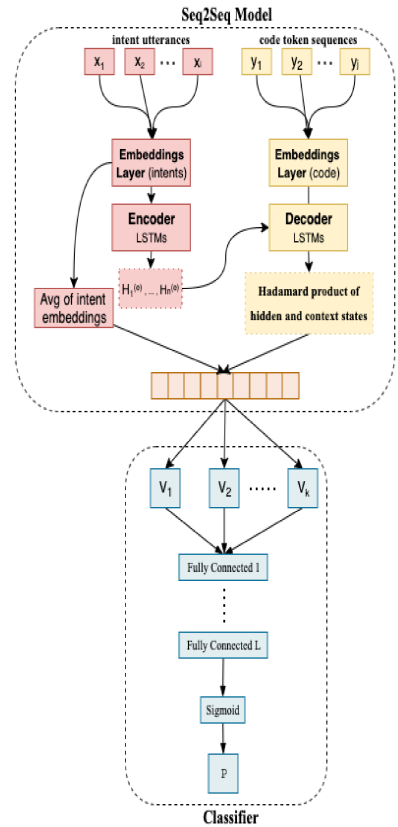

Figure 8: The baseline model of the hidden states of our project

Table 5: Binary Classifier Eval. Metrics for models.

| Binary Classifier | Loss | Accuracy (%) | AUC ROC | AUC PR | F1 |
| --- | --- | --- | --- | --- | --- |
| current GloVe | 0.69 | 51.60 | 0.51 | 0.53 | 0.68 |
| current Word2Vec | 0.69 | 51.50 | 0.52 | 0.55 | 0.68 |
| current GloVe + CSN | 0.69 | 51.50 | 0.44 | 0.47 | 0.68 |
| current Word2Vec + CSN | 0.69 | 51.50 | 0.50 | 0.51 | 0.68 |
| Hidden State Vectors | 0.69 | 51.60 | 0.51 | 0.65 | 0.68 |

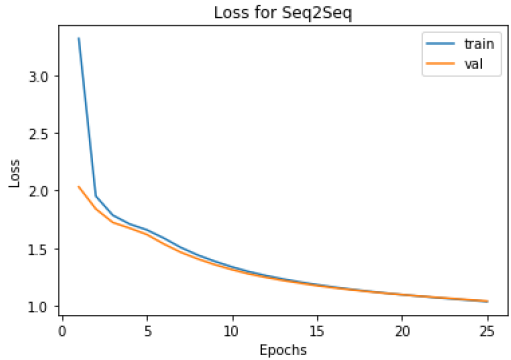

Figure 9: Current Glove Seq2Seq Loss

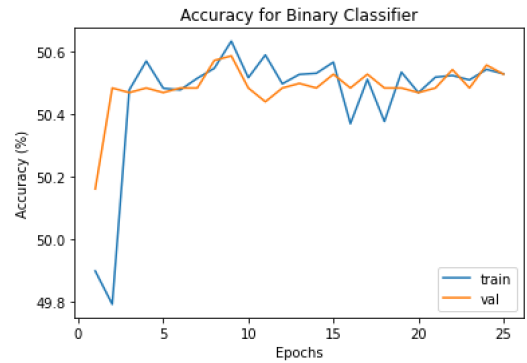

Figure 12: Hidden State Current Seq2Seq Accuracy

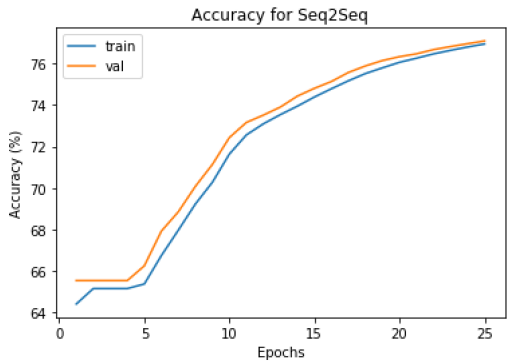

Figure 10: Current Glove Seq2Seq Accuracy

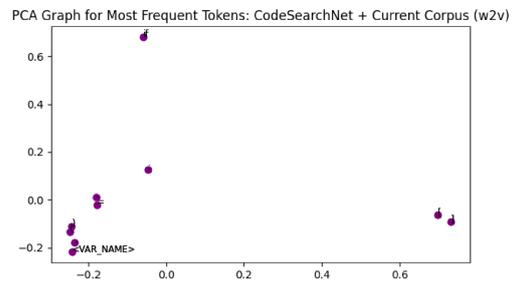

Figure 13: PCA Plot of Current + CSN Corpus for Word2Vec

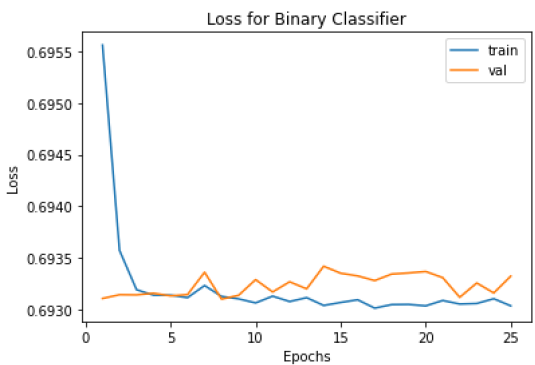

Figure 11: Hidden State Current Seq2Seq Loss

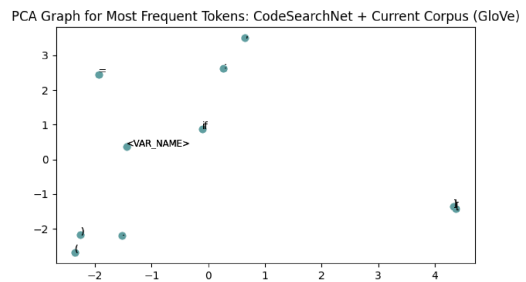

Figure 14: PCA Plot of Current + CSN Corpus for GloVe

# 6 Evaluation

## 6.1 Pre-trained Embedding VS Decoder Hidden States Vectors

Pre-trained embeddings are able to capture semantic information and learn similarities between other tokens. The results of our experiments have shown that pre-trained embeddings out-perform rather than using an elementwise product of hidden state vector and the context state vector as representations. As shown in our Seq2Seq (table 4) model, we observe that the hidden state vector model has the highest loss and the second worst accuracy. The GloVe and Word2Vec based embeddings for the code, using the current dataset, both our perform the hidden state vector model. An element wise operation on the hidden state and the context state however does not capture the semantics as effectively as how the the code based pre-training embeddings do.

## 6.2 Current corpus with or without development corpus

So, from the results, we can note that the seq2seq models are able to generalize slightly better with the pre-trained embeddings trained on the current corpus over the embeddings trained using the development corpus CSN in addition. We suppose this is because we find that the development corpus is probably not sufficient for helping the models learn the semantic embeddings of the tokens. We suppose that some of the code tokens may have had a different contextual meaning in general than in comparison to the local current corpus context. This might have been the main issue. Quantitatively (table 4), both the current corpus trained embeddings GloVe and W2V perform better than their development corpus based counterparts in terms of loss and accuracy of the test set. Clearly, the inclusion of a development corpus proved to be slightly inefficient in helping models capture the intricacies of the Python language tokens in understanding the contextual meaning for a given task. This situation is very much similar to a word based situation in a natural language: for example very analogous to the word "pass": probably meaning to pass an object in a generic object rather than the exam related pass term within an educational text corpus. Similarly, within a coding corpus maybe the token (*) might have been used for multiplication more often (probably because of the integer/numeric datatype dominance in the dataset) in the current corpus rather in the development corpus where it could have been used more as an operator for enclosing arguments within a method. Similarly, the use of some of these tokens may have a different meaning within a current corpus than in a generic context.

## 6.3 GloVe vs. Word2Vec

Word embeddings have are able to capture semantic information about text. We evaluate two different word embeddings GloVe and Word2Vec. In the Seq2Seq model, we see that GloVe outperforms Word2Vec by almost 2% in terms of accuracy. Also, in the Binary Classifier model we see that GloVe has marginal better accuracy (0.1%) but word AUC ROC and PR. Both word embeddings are able to capture similarities between tokens as shown in fig. 13 and fig. 14. Interestingly, Word2Vec maps similar tokens such as "(" and ")" closer together than GloVe. We can see that GloVe and Word2Vec improve the performance of the model with GloVe marginally outperforming Word2Vec.

# 7 Conclusion

Overall, we were able to implement a Seq2Seq model + binary classifier that is able to generate code snippets with a relatively high degree of accuracy from a given context (task), but however is only able to classify these produced snippets with the correct task intent with an accuracy a little bit above random chance. In order to execute our code generation and classification task, we combined and duplicated files from the CoNaLa (code natural language) and StaQC (Stack Overflow Question-Code pairs) datasets. Specifically, we combined training, testing, and manually mined samples from the CoNaLa dataset, with a portion of the StaQC dataset containing single code answer posts in the Python3 programming language. From these, we created our own dataset which consists partially of matched (positive samples) question IDs, task intents, and their corresponding code snippets, as well as mismatched (negative samples) task intents and code snippets. In the future, we want to continue doing experiment on our model, such as having a better trained embeddings, trying a combination of hidden states and the embeddings, implementing a better parsing technique for our model to understand a larger variety of code snippets.